# JANUS: Fast and Flexible Deep Learning via Symbolic Graph Execution of Imperative Programs*

Eunji Jeong
Seoul National University

Sungwoo Cho
Seoul National University

Gyeong-In Yu
Seoul National University

Joo Seong Jeong
Seoul National University

Dong-Jin Shin
Seoul National University

Byung-Gon Chun[†]
Seoul National University

## Abstract

The rapid evolution of deep neural networks is demanding deep learning (DL) frameworks not only to satisfy the requirement of quickly executing large computations, but also to support straightforward programming models for quickly implementing and experimenting with complex network structures. However, existing frameworks fail to excel in both departments simultaneously, leading to diverged efforts for optimizing performance and improving usability.

This paper presents JANUS, a system that combines the advantages from both sides by transparently converting an imperative DL program written in Python, the de-facto scripting language for DL, into an efficiently executable symbolic dataflow graph. JANUS can convert various dynamic features of Python, including dynamic control flow, dynamic types, and impure functions, into elements of a symbolic dataflow graph. Our experiments show that JANUS can achieve fast DL training by exploiting the techniques imposed by symbolic graph-based DL frameworks, while maintaining the simple and flexible programmability of imperative DL frameworks at the same time.

## 1 Introduction

In recent years, deep neural networks have been widely used in various application domains such as computer vision, speech, and natural language processing for their powerful capabilities of extracting abstract features from data. Scientists have created deep learning (DL) frameworks – TensorFlow [1], PyTorch [32], Caffe2 [12], MXNet [7], and many more [3, 13, 29, 31, 41, 43, 46, 49] – to improve the performance of deep neural networks in various jobs and promote the use of deep neural networks in both production and research.

Such DL frameworks can be classified into two distinct families depending on their execution models. One family comprises frameworks that base their execution on symbolic graphs constructed from DL programs. The other family consists of frameworks that directly execute DL programs in an imperative manner.

**Symbolic graph execution.** Frameworks such as TensorFlow [1], Caffe2 [12], and MXNet [7] formulate neural networks as symbolic dataflow graphs. Graph vertices denote the states and operations of a neural network, while graph edges indicate the flow of data between vertices. Operations in the graph are executed as their dependencies are solved, similar to how most dataflow systems process dataflow graphs [10, 18]. The graph representation allows the framework to identify which operations can be run in parallel, and apply various compiler optimization techniques such as common subexpression elimination or constant folding to generate optimized versions of graphs. Moreover, it is easy to process dataflow graphs on accelerator devices or deploy graphs across multiple machines by assigning an operation to the appropriate device or machine [25].

However, the separation of building a symbolic graph and executing it complicates user experience, because users are not actually running any numerical computations when defining neural networks through the framework interface. Rather, they are constructing graphs that will be executed later through separate functions.

**Imperative program execution.** In contrast, frameworks including PyTorch [32], TensorFlow Eager [41], and MXNet Imperative [28] have adopted the execution model of running operations imperatively, without going through a separate graph construction phase. Stemming from popular Python libraries for scientific, numerical computation such as NumPy [48] and Scikit-learn [5], this imperative approach is useful for rapidly experimenting and working with new neural network models, particularly those with complex structures. The native control flow statements of Python can be exploited to build mod-els of interest. Unfortunately, skipping the formation of a dataflow graph means that such frameworks lose the chance to apply the many optimizations that were possible in the symbolic graph execution model, leading to significant performance differences for certain models.

---

*Appeared in NSDI '19. [†]Corresponding author.

The different characteristics of DL frameworks suggest that we cannot achieve high performance and good usability at the same time. To reach high performance, we must sacrifice framework usability to a certain extent, and vice versa. Otherwise, users are forced to resort to an awkward approach of learning how to use several frameworks and switching between them according to the current task in hand.

**From imperative programs to symbolic graphs.** In this paper, we propose to transparently convert imperative Python DL programs into symbolic dataflow graphs before execution. By not altering the user-facing interface for building neural networks, we maintain the flexible programmability of frameworks with imperative execution models. At the same time, behind the scenes, we execute the symbolic graph versions of the imperative programs to enjoy the performance optimizations done by symbolic graph execution models.

However, this approach introduces a technical challenge of capturing the dynamic semantics of an imperative Python program in a symbolic dataflow graph. The dynamic aspects of Python, including dynamic control flow, dynamic typing, and impure functions, must be embedded in a symbolic graph correctly while providing the performance of symbolic graph execution frameworks.

To this end, we present JANUS, a DL framework that achieves the best of both worlds by receiving an imperative DL program as input and creating symbolic graphs of the program accordingly with speculative program context assumptions. JANUS makes environment assumptions on the program context (e.g., constant variables and branches) based on past iterations to simplify the dynamic nature of the program and transform the program into a symbolic graph. These assumptions are speculative, because the context may change during execution; an incorrect assumption results in an invalidation of a symbolic graph, in which case JANUS falls back to imperative execution to guarantee correctness. For design (Section 4.3.1) and implementation (Section 4.3.2) reasons, JANUS converts only the subset of Python programs into the efficient symbolic graphs, but the rest of them still can be executed imperatively, ensuring the full Python coverage.

We have implemented JANUS on TensorFlow 1.8.0 [1]. To demonstrate the performance of JANUS, we evaluated JANUS with 11 imperative DL programs in five categories: convolutional, recurrent, and recursive neural networks,

```
1   class RNNModel(object):
2     def __call__(self, sequence):
3       state = self.state
4       outputs = []
5       for item in sequence:
6         state = rnn_cell(state, item)
7         outputs += [state]
8       self.state = state
9       return compute_loss(outputs)
10
11  for sequence in sequences:
12    optimize(lambda: model(sequence))
```

Figure 1: A Python program that implements training process of a recurrent neural network (RNN) in an imperative manner. For each item in the sequence, rnn_cell function is called to produce the next state required for the next rnn_cell invocation. After finishing up processing the whole sequence, the model holds the final state by replacing self.state attribute for processing the next sequence.

generative adversarial networks, and deep reinforcement learning models that extensively use the dynamic features of Python. JANUS converted the programs into symbolic dataflow graphs successfully, trained the models to reach target accuracy with up to 18.7 times higher throughput compared to TensorFlow Eager, while executing the identical imperative programs.

## 2 Challenges and Proposed Solution

### 2.1 Challenges

Converting an imperative program written in Python into a DL dataflow graph brings on many challenges, because dataflow graphs consist of a restrictive set of operations, lacking the dynamic semantics of the programming language. More specifically, various characteristics of a Python program, such as the execution count and execution order of statements, the types of expressions, or the global program execution state, can only be determined after the program is actually executed. For the rest of this paper, we will refer to these characteristics as the *dynamic* features of Python. In contrast, DL dataflow graphs are expected to be defined before the computation starts, to apply aggressive graph optimizations and efficiently schedule the graph operations by viewing the entire graph. In this sense, DL dataflow graphs are usually considered to be *static* [24, 31, 32]. The difference in characteristics makes it difficult to embed dynamic Python features in static dataflow graphs.

Figure 1 depicts a DL program written in Python, of which semantics are difficult to be captured in a dataflow

| Frameworks | Imp. pgm | Sym. exec | Correctness | | | Optimization w/ runtime info | Language |
|---|---|---|---|---|---|---|---|
| | | | DCF | DT | IF | | |
| **Symbolic**: TensorFlow (TF), Caffe2, MXNet | × | ○ | – | – | – | – | Python |
| **Imperative**: PyTorch (PTH), TF Eager, DyNet | ○ | × | – | – | – | – | Python |
| **Imperative to Symbolic** | | | | | | | |
|   Tracing: TF `defun`, PTH JIT `trace`, MXNet Gluon | ○ | ○ | × | △ | × | ○ (unsafe) | Python |
|   JAX | ○ | ○ | ○ | △ | × | ○ (unsafe) | Python subset |
|   Swift for TensorFlow (S4TF) | ○ | ○ | ○ | – | ○ | × | Swift |
|   TF AutoGraph | ○ | ○ | △ | △ | △ | ○ (unsafe) | Python subset |
|   PTH JIT `script` | ○ | ○ | ○ | △ | △ | × | Python subset |
|   JANUS | ○ | ○ | ○ | ○ | ○ | ○ | Python |

Table 1: Comparison of DL frameworks with respect to correctly supported features for converting imperative programs into symbolic graphs ("Correctness") and the ability to optimize the generated graphs with the information given only at program runtime ("Optimization w/ runtime info"). Optimizations can be incorrect in some frameworks ("○ (unsafe)"), not preserving the original semantics of Python. The host language is also specified.

graph **correctly** due to the following representative dynamic features of Python.

- **Dynamic control flow (DCF)** Conditional branches and iterative loop constructs have different execution paths depending on intermediate values. Lines 5-7 of Figure 1 show an example of an iterative loop construct used in a DL program. Such control flow statements are intensively used in Python and must be correctly represented in the dataflow graph.

- **Dynamic types (DT)** Python is a dynamically-typed language, i.e., the type of a Python expression can only be determined at program execution time. The example program in Figure 1 does not have any type annotations (e.g. `int` or `float`), which makes it difficult to statically decide the type of target dataflow graph operations. Furthermore, various non-numerical types of Python, such as lists, dictionaries, and arbitrary class instances, are even harder to be converted into elements of a dataflow graph, of which vertices usually output numerical arrays.

- **Impure[1] functions (IF)** Another useful feature for using Python is the ease of accessing and mutating global states within functions. In Figure 1, the function `__call__` reads from and writes to an object attribute[2] at Lines 3 and 8, to pass the final state of a sequence to the next sequence. Since the modified global states can make the following function call behave differently, such reads and writes of global states must be handled correctly while generating dataflow graphs.

Moreover, correctness is not the only issue when converting an imperative program; achieving the high **performance** of state-of-the-art symbolic graph execution DL frameworks is also a challenge on its own. State-of-the-art frameworks require additional information on dynamic types and control flow in order to optimize graph execution. However, a naïve, one-shot converter would be unable to extract this information from an imperative program before execution, and thus is incapable of supplying frameworks with such hints. For instance, if the input `sequence` at Line 2 in Figure 1 is expected to always have a fixed length, then that information can be exploited to unroll the following loop at Line 5 when generating the corresponding dataflow graph. It is unclear how a naïve converter would do this without actually executing the program to check the loop length.

## 2.2 Related Works

Previous works that try to translate a Python DL program into a dataflow graph either fail to capture the important dynamic semantics of Python, or run in slower performance due to the lack of sufficient information at graph build time. Table 1 summarizes state-of-the-art DL frameworks alongside their execution models and their status regarding the coverage and efficiency of graph conversion support.

Tracing-based graph generation approaches such as PyTorch's JIT compiler (`torch.jit.trace`) [32], MXNet Gluon [29], and the `defun` [44] functionality of TensorFlow Eager [41] execute the imperative program once, and convert the single execution trace directly into a dataflow graph. Though this approach enables generating optimized symbolic graphs with sufficient information gathered from a specific execution trace, it fails to capture dynamic semantics of the Python interpreter correctly, leading to incorrect computation results for dynamically changing execution paths, dynamic types of non-tensor or non-input expressions, or impure functions of Python at runtime. Moreover, these approaches currently do not give any feedback about incorrectly-converted control flows to users, making the problem even worse.

---
[1] A *pure function* is a function whose return value is determined only by its parameters, and has no side effects.
[2] "class members" in C++ terminology, except that the attributes are stored in dictionaries, without fixed data layout.

On the other hand, there exist other approaches that select a less-dynamic host language and therefore succeed in capturing the wider semantics of source programs. JAX [13] limits the Python syntax and supports converting only pure-and-statically-composed functions. S4TF [43] supports Swift, losing the merit of supporting Python, the de-facto standard programming language for DL programming, and introduces new programming models that most DL researchers are unfamiliar with. Moreover, since the graph conversion occurs before actually executing the program, these approaches can miss the opportunity to further optimize the graph with the information only obtainable during the program execution. For example, always converting a Python loop into control flow operations can be sub-optimal if the loop iteration count is known to be fixed.

Concurrent works including AutoGraph-enabled TensorFlow `defun` functionality [27] and the "scripting" mode of PyTorch JIT (`torch.jit.script`) [32] also have limitations. AutoGraph makes users to explicitly provide the necessary information, or generates incorrect or sub-optimal graph in some cases, all of which could be avoided if sufficient information existed. For example, users must explicitly specify the types of Python lists, prohibiting the dynamic typed or heterogeneous elements. For another example, for dynamic control flow statements, the statements with non-tensor predicates are always unrolled, which is error-prone, and the statements with tensor-typed predicates are always converted to control flow operations, which can be sub-optimal. In the "scripting" mode of PyTorch JIT, users must use TorchScript, a subset of Python which does not allow variables to have dynamic types. Further graph optimizations based on the runtime information are also not possible.

### 2.3 Proposed Solution: Speculative Graph Generation and Execution

Existing optimizers and compilers for dynamic languages suggest a useful technique for performing such conversions from imperative programs to symbolic dataflow graphs: *speculative optimization*. Managed language runtimes have succeeded in exploiting the inherent static nature of dynamic programs which rarely changes during the execution to convert them into static, low-level representations while maintaining correctness. For example, JavaScript just-in-time (JIT) compilers convert dynamic JavaScript programs into efficient machine code, and this conversion is done speculatively assuming that the program inherently maintains some statically fixed structures over repeated executions. In case this assumption breaks, the program falls back to the interpreter and attempts to compile the program again with different assumptions.

We propose to adopt this concept of speculative optimization when converting imperative DL programs into symbolic dataflow graphs. Converting various dynamic features like dynamic control flow and impure functions correctly may impose some inevitable overheads if we generate dataflow graphs in a conservative manner. To overcome this challenge, JANUS makes assumptions about the program's behavior based on the runtime profiling information, and generates a symbolic graph tailored for the assumptions. This speculatively constructed dataflow graph can show much better performance compared to the conservative counterpart due to specializations. If the assumptions do not hold, JANUS builds a new dataflow graph based on different assumptions. Since a DL program comprises a number of iterations of an optimization procedure, the speculative approach is a good fit since the interpreter is likely to execute specific code blocks of the program repeatedly.

Unlike the JIT compilers of managed language runtimes, however, the goal of JANUS is not to optimize the host language execution itself. In fact, when running imperative DL programs, the execution time of the language runtime is usually much shorter compared to the execution time of the mathematical operations for DL, such as convolution or matrix multiplication. However, since these mathematical operations are usually implemented in separate low-level language like C++, existing JIT compilers of managed language runtimes would execute them just as separated function invocations. Under such an execution model, it is impossible to see the multiple mathematical operations at once and apply compiler optimizations or execute them in parallel. On the other hand, JANUS understands the function invocations for such mathematical operations, and converts them into appropriate target graph operations, which can be optimized and be executed efficiently by symbolic graph executors.

## 3 JANUS System Design

In this section, we introduce JANUS, a DL framework that receives an imperative DL program and either executes it as is directly, or generates a symbolic graph version of the program and executes the graph instead.

The input program for JANUS is assumed to be written using the API and the programming model of existing imperative DL frameworks like TensorFlow Eager [41].

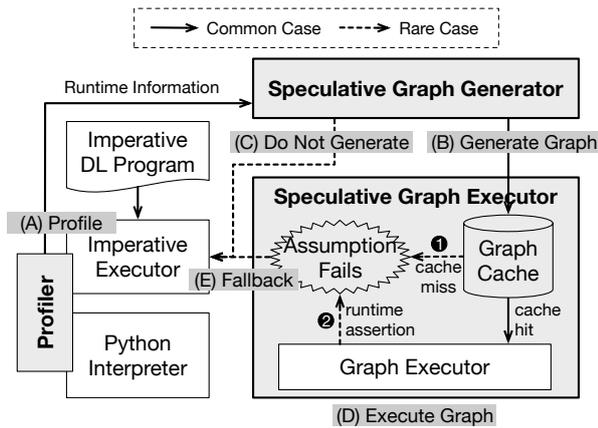

Figure 2: An illustration of the execution model of JANUS, showing how a DL program is processed by several components. *Profiler* observes imperative program execution and collects information to make the realistic assumptions. *Speculative Graph Generator* generates dataflow graphs from the program and hands the optimized graphs over to Speculative Graph Executor. The *Speculative Graph Executor* actually runs the generated graph and handles assumption failures.

Given an input program, JANUS extracts the main neural network computation part, over which the automatic differentiation is performed, and starts the speculative graph generation and execution process. From the user's point of view, the whole graph conversion and execution process is done transparently; in other words, the given DL program is automatically transformed into a corresponding graph representation without any interactions.

Figure 2 depicts the system components and the overall execution model of JANUS. The common case in which an efficient dataflow graph is utilized is depicted as solid lines in the figure, while the rare case where the graph representation is not available is depicted as dotted lines.

## 3.1 Fast Path for Common Cases

**Runtime profiling.** Once JANUS receives a DL program, the program is first executed imperatively, while the *Profiler* gathers runtime information required for making reasonable assumptions (Figure 2 (A)). Various information is collected, including control flow decisions on conditional branches, loop iteration counts for iterative loop constructs, variable type information, non-local variables, object attributes, and so on.

**Symbolic graph generation.** After a sufficient amount of information has been collected, the *Speculative Graph Generator* tries to convert the program into a symbolic dataflow graph with the assumptions based on the runtime information (Figure 2 (B)). To avoid making any hasty generalizations, JANUS does not begin graph generation until the executor has profiled the program for a certain amount of iterations.[3] First, JANUS traverses the abstract syntax tree (AST) of the DL program and generates the corresponding graph elements for each AST node, along with assertion operations that can validate the context assumption at runtime. Since JANUS targets DL programs, operations for automatic differentiation and model parameter updates are also automatically inserted if necessary. Next, the generated graph is further optimized by the post-processor, of which optimizations were not applicable to the original imperative DL program. Finally, the optimized graph and the assumption that were used to generate the graph are saved into the graph cache.

**Graph execution.** If a graph representation with correct assumptions regarding the program context is available, the *Speculative Graph Executor* executes the symbolic graph (Figure 2 (D)). Note that the same graph can be reused multiple times, given that the runtime context assumption holds for future invocations.

## 3.2 Accurate Path for Rare Cases

**Assumption failure.** Handling the assumptions is important to guarantee the correctness of the converted graph. If an assumption is proven to be wrong, the associated graph cannot be executed for the current runtime as it may produce incorrect results. Instead, JANUS falls back to the imperative executor (Figure 2 (E)) and resumes runtime profiling to make more relaxed assumptions for subsequent executions.

Assumptions that can be validated before actually executing the associated graph, such as type assumptions on input arguments, are checked when retrieving the graph from the graph cache (Figure 2 ①). In the unfortunate case where such an assumption is wrong, JANUS regards this as a cache miss and falls back to imperative execution.

On the other hand, for assumptions that can only be validated during graph execution (Figure 2 ②), it can be erroneous to simply abort the current execution to fall back to the imperative executor, because the global state may have been changed during the current execution. To solve this issue, JANUS defers state update operations until every assumption is validated (Section 4.2.3). This way, even if an assumption turns out to be wrong during computation, no state update operation has been triggered yet and thus no state has been mutated. Knowing this, the system can safely stop the current execution. In other words, states are updated in an all-or-nothing manner.

---

[3]We found that 3 iterations were enough to come up with a decent program context assumption, for our experimental workloads.

In order to validate an assumption, a runtime assertion is encoded into the symbolic graph as an operation called `AssertOp`. The `AssertOp` aborts the graph execution if the given condition fails. It also reports which assumption has been broken, and this information is used to give up further optimizations that rely on the assumptions that repeatedly break.

**Imperatively executed programs.** With Turing-complete graph representations, any Python program can be represented as a symbolic graph, in theory. However, the *Speculative Graph Generator* does not convert every single Python feature into a symbolic graph operation (Figure 2 (C)). For example, to ensure the all-or-nothing characteristic of state updates, programs that include invisible state mutations are not converted into symbolic graphs. Some complicated Python features such as *coroutine*s and *generator*s are also not converted, since they do not have any clear graph representations. Section 4.3 describes the design choices and current limitations of the *Speculative Graph Generator* in terms of Python coverage. In spite of such limitations of the *Speculative Graph Generator*, however, it is worth noting that JANUS users can still freely use the all features of Python on the imperative executor.

## 4 Symbolic Graph Generation

In this section, we describe in detail how JANUS converts an imperative DL program into a symbolic dataflow graph. We start the section by showing the conversion process of a basic DL program free of dynamic features (Section 4.1). Next, we explain how JANUS converts dynamic features of Python, including dynamic control flow, dynamic types, and impure functions, into symbolic graph operations (Section 4.2). JANUS uses the runtime information to simplify the dynamic program and treat it as a program of only static aspects, which is then easily transformed into a static graph. Finally, we discuss the Python coverage limitations of the *Symbolic Graph Generator* (Section 4.3). More thorough discussion about the Python coverage of JANUS is in Appendix A.

For simplicity, we describe our design using various operations of TensorFlow [1], a widely-used DL framework. However, our design is not necessarily coupled with TensorFlow and can be applied to other DL frameworks.

### 4.1 Graph Generation Basics

Figure 3(a) is a simple, imperative Python program that calculates a linear model, written as a pure function without any dynamic control flow or arbitrary Python objects. We use this program as an example to show the basic graph conversion process.

```
1  def loss_fn(x, y):
2      y_ = 0.5 * x + 1.5
3      return (y_ - y) ** 2
```

(a) Source code of a DL program calculating a linear model

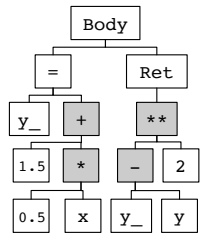
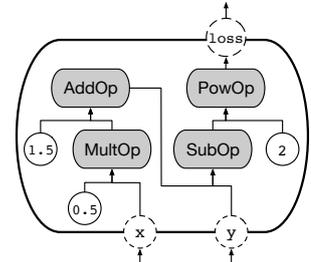

(b) AST of `loss_fn`  (c) Generated graph from `loss_fn`

Figure 3: The Python source code, AST, and symbolic graph of a simple linear model that receives several external inputs. The static features of the program are represented as nodes in the AST, which in turn are converted to vertices of the symbolic graph.

Input parameters (`x` and `y`) are converted into graph input objects that require external inputs in order to execute the graph. In the case of TensorFlow, this corresponds to `PlaceholderOp`[4]s. At runtime, they are filled with the actual argument values. The return value of the `return` statement is marked as the computation target of the graph, so that we can retrieve the value after executing the graph.

Python literals such as `0.5`, `1.5` and `2` are simply converted into operations that output constant values – `ConstantOp` for TensorFlow. The conversion of mathematical operators is done by finding the corresponding mathematical graph operations and replacing them one-to-one. For standard Python operators such as `+` and `**`, JANUS places the appropriate primitive calculation operations in the graph, like `AddOp` and `PowOp` for TensorFlow.

An assignment to a Python local variable and a value retrieval from the same variable is converted into a connection between two operations, just as in Pydron [30]. Figures 3(b) and 3(c) illustrate how such a connection is made for the variable `y_` in Figure 3(a), along with the rest of the program.

### 4.2 Dynamic Features

In addition to the basic features, JANUS converts the dynamic features of Python into the elements of the symbolic DL graph as well to provide the performance of dataflow graphs while maintaining the same programmability of imperative DL frameworks. Moreover, JANUS exploits the fact that the dynamism in Python DL programs can often be simplified to static dataflow, treating a dynamic program as a program of only static aspects with appropriate program context assumptions. Context assumptions are generated based on the profile information JANUS gathers at runtime.

---
[4]`PlaceholderOp`s are unique operations that generate errors unless they are provided with external inputs before graph execution. TensorFlow expects users to feed a dictionary `{ph1: v1, ph2: v2, ...}` to a `PlaceHolderOp`.

### 4.2.1 Dynamic Control Flow

**Basic translation rules.** Among various dynamic control flow statements, JANUS focuses on conditional branches, loop constructs, and function calls, similar to Pydron [30]. As shown in Pydron, these three constructs are enough to express most complex dynamic control flows in Python. Furthermore, they can all be expressed using special control flow graph operations proposed in recent works [20, 50] as follows.

Python's conditional statement, the `if` statement, can be obtained by combining *switch* and *merge* primitives. The switch and merge primitives, originating from classic dataflow architectures [2, 9, 11], act as demultiplexers and multiplexers, respectively, selecting a single path to pass their inputs or outputs. In TensorFlow, the `SwitchOp` and `MergeOp` [50] serve as symbolic dataflow graph counterparts for these primitives, allowing JANUS to plant conditional branches in graphs.

The iterative statements of Python, `while` and `for`, are handled by using the switch and merge primitives together with loop context primitives that hold iteration *frame*s. TensorFlow conveniently provides `EnterOp`, `ExitOp`, and `NextIterationOp` [50] for creating iteration frames and passing values over them.

Finally, for function calls, a separate graph is generated for the callee function, and a function invocation operation that points to the generated graph is inserted in the position of the function calls. Recent work proposes a TensorFlow implementation of this operation called `InvokeOp` [20], which can represent an invocation of a recursive function with automatic differentiation support.

**Speculative graph generation: unrolling and inlining.** If JANUS detects that only a single particular path is taken for a certain control flow statement during profiling, JANUS presumes that the control flow decision is actually fixed. The system replaces the control flow operation with an assertion operation that double-checks the assumption for this control flow decision, and proceeds with graph generation as if the control flow statement were unrolled. This allows JANUS to remove control flow operation overheads and apply graph optimizations such as common subexpression elimination or constant folding in broader portions of the graph. If the assertion operation fails, JANUS falls back to imperative execution.

To be more specific, for conditional branches, if the program takes only one side of the branch during profiling, JANUS generates that particular side of the branch in the final graph without any switch or merge primitives and adds an assertion operation that can detect a jump to the other side of the branch. For iterative statements, if the number of iterations of a loop is discovered to be fixed, JANUS unrolls the loop with this fixed iteration count, and adds an assertion operation to check that the number of iterations is indeed correct.

For function calls, if the callee is expected to be fixed for a function call at a certain position, JANUS inlines the callee function body inside the caller unless that function call is identified as a recursive one. In addition, for callee functions whose implementation is already known for JANUS, e.g., the functions provided by the framework such as `matmul()` or `conv2d()`, or Python built-in functions like `print()` or `len()`, JANUS adds the corresponding graph operations which behave the same as the original callee functions, based on the prior knowledge about their behaviors. Section 4.3.1 includes more details and limitations about such function calls.

### 4.2.2 Dynamic Type

**Basic translation rules.** The types of all expressions within a Python program must be known before JANUS can convert the program into a symbolic graph, because graph operations require operands to have fixed types. This is a challenging task for Python programs because we cannot determine the type of an arbitrary Python expression before actually executing the expression. Fortunately, it is possible to infer the types of some expressions, given the types of other expressions; for example, it is clear that the variable `c` in `c = a + b` is an integer if `a` and `b` are integers.

As a basic rule, JANUS converts numerical Python values such as scalars, list of numbers, and NumPy [48] arrays into corresponding tensors, and converts non-numerical values, including arbitrary class instances, into integer-typed scalar tensors which hold pointers to the corresponding Python values. Next, JANUS infers the types of other expressions that are derived from expressions covered by the basic rule.

**Speculative graph generation: specialization.** Expressions whose types cannot be inferred from other expressions require a different measure. For instance, it is impossible to identify the types of input parameters for functions, or Python object attribute accesses (`obj.attr`) without any external clues. Similarly, inferring the return types of recursive function calls is also challenging due to the circular dependencies. To make proper assumptions about the types of such expressions, *Profiler* observes the types of the expressions during imperative executions. Given these context assumptions, JANUS can finish inferring the types of remaining expressions, and construct a specialized dataflow graph accordingly.

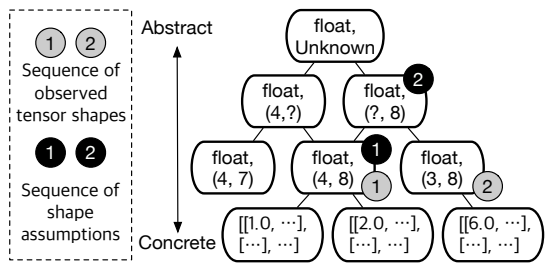

Figure 4: Type, shape, and value specialization hierarchy for an example tensor.

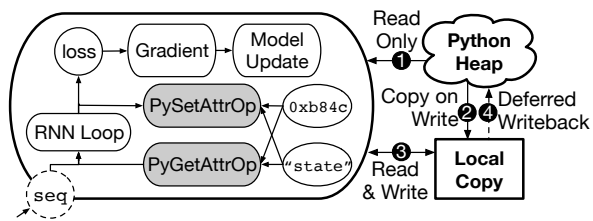

Figure 5: Symbolic dataflow graph generated graph from Figure 1 and the global states.

In addition, JANUS makes further assumptions about the expressions to apply more aggressive optimizations. For numerical expressions, we can try to specialize the shape of tensors before constructing the graph. Furthermore, if a Python expression always evaluates to the same value while profiling, JANUS converts it into a constant node in the dataflow graph. With statically determined shapes or values, the graph can be further optimized, or even be compiled to the efficient machine code [45].

Figure 4 shows an example hierarchy of shapes and values that a certain tensor may have. After profiling the first few runs, JANUS finds out that even though the values of the tensor are different every time, they all have the same shape, for example (4, 8), as in the figure. JANUS exploits this information to generate a dataflow graph with an assumption that the shape of this tensor is (4, 8). When the assumption fails, JANUS tries to relax the assumption. For instance, in case the tensor has a shape (3, 8) for the next iteration to process a different size of mini-batch, JANUS modifies the assumption to suit both shapes (4, 8) and (3, 8), resulting in another dataflow graph with a shape assumption of (?, 8). The system does not have to repeat the graph generation process for a possible future case in which the example tensor has yet another unpredicted shape of (2, 8) or (6, 8).

#### 4.2.3 Impure Functions

**Naïve translation rules.** It is common for a Python function to access global variables to calculate return values and have side-effects, mutating its enclosing Python context during execution. Likewise, it is common for a Python DL program to read from and write to global states such as global or nonlocal variables and heap objects. JANUS respects this characteristic and handles global state accesses alongside symbolic graph execution.

A trivial solution is to use TensorFlow's `PyFuncOps`, which can execute arbitrary Python functions as graph operations. A function for reading and updating a certain global state can be created and inserted in the appropriate position within the graph. However, this trivial approach has clear limitations. First, since only one Python function can be executed at a time due to the global interpreter lock (GIL), the overall performance can be reduced when multiple operations should be executed in parallel. It also complicates the fallback mechanism of JANUS. If a global state has already been mutated before the fallback occurs, instead of starting the imperative executor from the function entrance at fallback, execution must start from the middle of the function to be correct, by mapping the state update operation with corresponding Python bytecode.

**Optimized graph generation: deferred state update.** To make things simpler and also faster, JANUS does not mutate global states in place on the fly. JANUS instead creates local copies of global states, and mutates only the local copies during symbolic graph execution.

Figure 5 shows the symbolic dataflow graph version of the program in Figure 1, which includes the object attribute expressions (`self.state`) that access and mutate the global states. We add new graph operations `PyGetAttrOp` and `PySetAttrOp` to represent Python attribute read and write. Each of them receives an object pointer (`0xb84c`) and a name of the attribute (`"state"`) as inputs, and behaves as follows: ① The `PyGetAttrOp` can access the Python heap to read the state unless a corresponding local copy exists. ② When the `PySetAttrOp` wants to update the attribute, a new value is inserted to the local copy instead of directly updating the Python heap. ③ Further read and write operations are redirected to the local copies. Note that JANUS inserts appropriate dependencies between `PyGetAttrOps` and `PySetAttrOps` if necessary to prevent any data hazards. ④ After the graph executor finishes this run, the local copies are written back to the Python heap. Global or nonlocal variables can also be regarded as the object attributes, where the global variables are the attributes of the global object, and the nonlocal variables are the attributes of the function's closure objects. Subscript expressions (`obj[subscr]`) are similarly implemented with equivalent custom operations, `PyGetSubscrOp` and `PySetSubscrOp`.

By not mutating the Python heap directly, JANUS can always bypass the Python GIL to execute more read and write operations in parallel. In addition, the fallback mechanism of JANUS can be simplified thanks to the all-or-nothing based state update mechanism.

### 4.3 Imperative-Only Features

Albeit being able to support a wide range of imperative DL programs, the current JANUS graph generator does not convert some particular features of Python into dataflow graph elements. Programs with such features are executed only on the imperative executor.

#### 4.3.1 Coverage Limitations from Design

**Alignment with the design principles.** To be aligned with the design of JANUS in previous sections, the JANUS graph generator does not convert some features of Python. For example, to keep the implementation of local copies of global state simple (Section 4.2.3), Python objects with custom accessor functions (e.g., `__setattr__`) are not supported by the JANUS graph generator. Also, a function should always return the same type of value, to infer the type of call expressions (Section 4.2.2).

**External function calls.** JANUS must understand the behavior of the external functions, i.e., the framework-provided functions or foreign functions[5], to convert them into corresponding graph operations. The JANUS graph generator converts the external functions into the graph operations based on a separate whitelist. Most of the framework-provided functions such as `matmul` or `conv2d`, and many commonly-used Python built-in functions such as `print` or `len` are included in this whitelist. We plan to cover more functions in the Python standard library.

JANUS handles such external functions with extra caution to ensure correctness. First, since the underlying assumption here is that the implementation of external functions never changes, JANUS prohibits the modification of the functions included in the whitelist. Also, if an external function includes state mutation (e.g., `assign()` in TensorFlow), the execution of the corresponding graph operation is deferred until all the other assumptions are validated, under the same principle about the deferred state update in Section 4.2.3.

#### 4.3.2 Coverage Limitations from Implementation

Currently, JANUS does not cover a few features from Python that do not have clear graph representations. Such Python features include *coroutine*s, *generator*s, in-line

---
[5]functions written in the languages other than Python

class definitions and in-line import statements. We plan to support these features as future work.

## 5 Implementation

We implemented JANUS on top of TensorFlow [1] 1.8.0 and CPython [33] 3.5.2. JANUS exploits the existing TensorFlow graph executor and TensorFlow Eager imperative executor as its components. In this section, we explain the modifications to existing systems, and then describe how JANUS supports data-parallel training.

**Modifications to existing systems.** TensorFlow has been modified for several reasons. First, to transparently separate out the neural network computation from the rest of the Python program without extra user intervention, the automatic differentiation functionality of TensorFlow Eager is modified to trigger JANUS graph conversion. Second, to share the model parameters between eager mode and graph mode, JANUS slightly modifies the parameter storing mechanism of TensorFlow Eager. Third, several custom operations had been added, including the `InvokeOp` and `PyAttrOp` as described in earlier sections.

CPython has also been modified to have bytecode-level instrumentation functionality for non-intrusive profiling. Without modifying the interpreter, instrumentation for the profiling should exist at the Python source-code level, which would significantly affect the performance and the debuggability of the imperative execution.

**Data-parallelization on JANUS.** Using multiple machines equipped with multiple GPUs is a common approach for accelerating DL jobs. We integrate JANUS with Horovod [36], a distributed training module for TensorFlow that encapsulates the MPI collective communication [15] (e.g. AllReduce and AllGather) as an operation inside the symbolic graph. After converting an imperative program into a dataflow graph, JANUS inserts appropriate communication operations to the graph in order to get the average of gradients generated by multiple workers. Since the generated dataflow graph contains both communication and computation operations, we can parallelize their execution and therefore achieve higher throughput.

## 6 Evaluation

We present experimental results that show how imperative DL programs can be executed both correctly and efficiently when converted into symbolic graphs on JANUS.

### 6.1 Experimental Setup

**Frameworks.** As baseline frameworks representing symbolic graph execution frameworks and imperative execution frameworks respectively, we use TensorFlow [1] and TensorFlow Eager [41]. We could run the same DL program on JANUS as on TensorFlow Eager, thanks to the transparent graph conversion feature of JANUS. In addition, to demonstrate the correctness of graph conversion of JANUS, we also compare JANUS with TensorFlow defun [44], which implements a trace-based graph conversion mechanism. TensorFlow-based frameworks have been chosen to avoid implementation-dependent performance differences.

**Applications.** We have evaluated JANUS with 11 models in five major neural network types, covering three convolutional neural networks (*CNN*; LeNet [22], ResNet50 [16], Inception-v3 [39]), two recurrent neural networks (*RNN*; LSTM [51], LM [21]), two recursive neural networks (*TreeNN*; TreeRNN [37], TreeLSTM [40]), two deep reinforcement learning models (*DRL*; A3C [26], PPO [35]), and two generative adversarial networks (*GAN*; AN [14], pix2pix [19]) as shown in Table 2. The datasets and the mini-batch sizes used for evaluation are also specified in the table.

These models are implemented in an imperative programming style, using a number of dynamic features in Python as shown in Table 2. First, large CNN models such as ResNet50 and Inception-v3 have conditional statements for handling batch normalization [17], which make them behave differently under particular conditions when training and evaluating the model. Next, RNNs include Python `for` loops, and they also include global state mutation statements to retain hidden states inside the models. Next, TreeNNs[6] require all three kinds of dynamic features. They include recursive function calls, and conditional statements to separate recursion base cases and inductive cases. They also include values with undecided type; the return type of a recursive function is unknown until the function returns certain values. In addition, they include the Python object access to fetch the information of the current subtree. For DRL models[7], Python `for` loops are used for handling an arbitrary length of the states of an episode, and global state mutation statements are used for storing the intermediate computation results to monitor the progress of the training. GAN models also use global state mutation statements for the same reason. All models use Python function calls, including Python class methods of high-level DL programming APIs such as Keras [8]. Training data instances fed into each neural network have different shapes over different training iterations, when the length of the dataset cannot be divided by the batch size.

---
[6]The implementation of TreeNN models on TensorFlow follows the recursion-based implementation with `InvokeOp` [20], and JANUS converts an imperative Python program into similar recursion-based graphs.
[7]The DL framework only handles model training and policy evaluation, and the environment simulation is handled by an external library [4].

| Category | Model | DataSet | BS | DCF | DT | IF |
|---|---|---|---|---|---|---|
| CNN | LeNet | MNIST [23] | 50 | × | ○ | × |
|  | ResNet50 | ImageNet [34] | 64 | ○ | ○ | × |
|  | Inception-v3 | ImageNet [34] | 64 | ○ | ○ | × |
| RNN | LSTM | PTB [51] | 20 | ○ | ○ | ○ |
|  | LM | 1B [6] | 256 | ○ | ○ | ○ |
| TreeNN | TreeRNN | SST [38] | 25 | ○ | ○ | ○ |
|  | TreeLSTM | SST [38] | 25 | ○ | ○ | ○ |
| DRL | A3C | CartPole [4] | 20 | ○ | ○ | ○ |
|  | PPO | Pong [4] | 256 | × | ○ | ○ |
| GAN | AN | MNIST [23] | 128 | × | ○ | ○ |
|  | pix2pix | Facades [47] | 1 | × | ○ | ○ |

Table 2: Categories, models, datasets, batch sizes ("BS"), and the dynamic features of the applications used for evaluation.

**Environments.** A homogeneous GPU cluster of 6 machines, connected via Mellanox ConnectX-4 cards with 100Gbps InfiniBand is used for evaluation. Each machine is equipped with two 18-core Intel Xeon E5-2695 @ 2.10 GHz, and 6 NVIDIA TITAN Xp GPU cards. Ubuntu 16.04, Horovod 0.12.1, CUDA 9.0, cuDNN 7, OpenMPI v3.0.0, and NCCL v2.1 are installed for each machine.

LeNet, LSTM, AN, and pix2pix models are evaluated on a single GPU, since these models and the datasets are regarded to be too small to amortize the communication cost of parallel execution. Similarly, TreeRNN, TreeLSTM, and A3C models are evaluated on CPUs on a single machine, since these models and datasets are regarded to be too small to amortize the communication between CPU and GPU. The other models are evaluated using multiple GPUs. ResNet50 and Inception-v3 models are evaluated using up to 36 GPUs, and LM is evaluated on up to 12 GPUs. The network bandwidth made the throughput of LM saturated on more than 2 machines with MPI collective communication, due to the huge parameter size of LM (0.83 billion parameters). Therefore, model convergence of LM is experimented with 6 GPUs. We evaluated the model convergence of PPO using 4 GPUs on a single machine, since the number of parallel actors used in the original paper was only 8.

## 6.2 Model Convergence

Figure 6 shows how the neural networks converge on various underlying frameworks, with ResNet50 with the ImageNet dataset, LM with the 1B dataset, TreeLSTM with the SST dataset, PPO with the Pong-v4 environment, and AN with the Facades dataset on four frameworks. For all evaluated models, JANUS, TensorFlow, and TensorFlow Eager succeeded to make the neural networks converge correctly as reported in literatures: 23.7% top-1 error for ResNet50 after 90 epochs, perplexity 47.5 for LM after 5 epochs, 82.0% binary accuracy for TreeLSTM after 4 epochs, 20.7 mean final score for PPO after 40M game frames, and 3.52 discriminator loss for AN after 30 epochs.[8] Also, JANUS could make the model to converge up to 18.7 times faster than TensorFlow Eager, while executing the identical imperative program. The performance difference between JANUS and TensorFlow was within 4.0%.

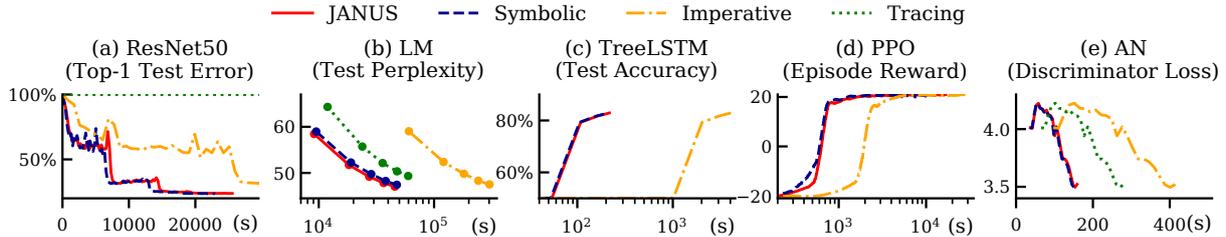

Figure 6: (a) The test error of ResNet50, (b) validation perplexity of LM, (c) test accuracy of TreeLSTM, (d) episode reward of PPO, and (e) discriminator loss of AN measured on JANUS, TensorFlow (Symbolic), TensorFlow Eager (Imperative), and TensorFlow `defun` (Tracing) according to the elapsed time in seconds. Each marker in (b) represents each training epoch, describing that per-epoch convergence is slower on TensorFlow `defun` compared to other frameworks.

On the other hand, trace-based TensorFlow `defun` failed to make the models to converge correctly. The ResNet50 model includes the conditional statement to distinguish the behavior of the batch-normalization [17] layer on model training and evaluation. If a user evaluates the initial accuracy before training the model by manipulating the model object attribute, TensorFlow `defun` converts the first execution trace into graph operations, which silently leads to an inaccurate result. Similarly, the LM model does not converge properly with TensorFlow `defun`, since it failed to capture state passing across sequences, due to its trace-based conversion mechanism. The TreeLSTM model could not be converted into the symbolic graph at all with TensorFlow `defun`, since it does not support recursive function call. We could not get the convergence metrics for PPO model with TensorFlow `defun`, as it does not support global state update statements. TensorFlow Eager converges slowly, since its training throughput is much lower than TensorFlow and JANUS. We next analyze the training throughput of the frameworks, excluding TensorFlow `defun`, which fails to make models converge correctly.

## 6.3 Training Throughput

### 6.3.1 Single-machine Throughput

Table 3 presents the training throughput of all models executed with JANUS, TensorFlow Eager, and TensorFlow on a single machine with a single GPU. As shown in the

[8]We measured the training loss with the official implementation in Tensorflow Eager [42].

| Model | (A) Imp. | (B) JANUS | (C) Sym. | $\frac{(B)}{(A)}$ | $\frac{(B)}{(C)}-1$ |
|---|---|---|---|---|---|
| LeNet | 7.94k | 25.84k | 26.82k | 3.25x | -3.6% |
| ResNet50 | 188.46 | 200.37 | 207.39 | 1.06x | -3.4% |
| Inception-v3 | 108.36 | 119.32 | 124.33 | 1.10x | -4.0% |
| LSTM | 2.75k | 22.06k | 22.58k | 8.03x | -2.3% |
| LM | 19.02k | 40.18k | 40.45k | 2.11x | -0.7% |
| TreeRNN | 20.76 | 988.72 | 928.66 | 47.6x | +6.5% |
| TreeLSTM | 7.51 | 138.12 | 141.71 | 18.4x | -2.5% |
| A3C | 220.66 | 1132.9 | 1178.6 | 5.13x | -3.9% |
| PPO | 596.80 | 1301.0 | 1306.4 | 2.18x | -0.4% |
| AN | 4.34k | 11.33k | 11.56k | 2.61x | -2.1% |
| pix2pix | 4.04 | 8.69 | 8.88 | 2.15x | -2.1% |

Table 3: Training throughput of all models evaluated on a single machine with a single GPU in JANUS, TensorFlow (Sym.), and TensorFlow Eager (Imp.). The numbers represent processed images/s for CNN and GAN models, processed words/s for RNN models, processed sentences/s for TreeNN models, and processed frames/s for DRL models.

table, JANUS outperforms TensorFlow Eager (imperative execution) by up to 47.6 times, and shows throughput similar to TensorFlow (symbolic graph execution) by up to 4.0% performance degradation. JANUS even performs slightly better (+6.5%) for TreeRNN, since there is no need to pre-process the input sentences, which are the tree-structured Python objects.

JANUS achieves bigger performance gains on RNNs, TreeNNs, DRLs, and GANs than on CNNs, since those networks have many concurrently executable operations. In addition, the performance gain of JANUS on a single machine is larger on models with fine-grained graph operations such as LeNet, LSTM, TreeRNN, A3C, and AN, compared to the models with coarse-grained operations such as ResNet50, Inception-v3, LM, PPO, and pix2pix, since the gain from bypassing the Python interpreter and applying compiler optimizations is bigger when the computation time of each operation is short.

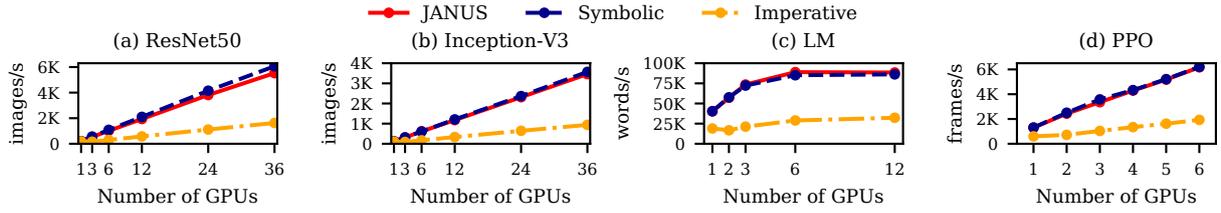

Figure 8: Training throughput for the ResNet50, Inception-v3, LM, and PPO models on JANUS, TensorFlow (Symbolic), TensorFlow Eager (Imperative), using varying numbers of GPUs.

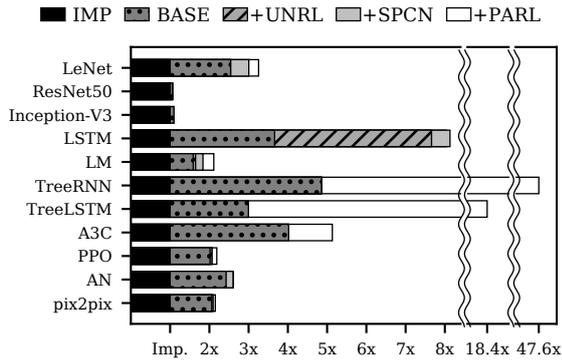

Figure 7: The contribution of optimizations to improve training throughput. Optimizations are cumulative. **+PARL** is the default configuration of JANUS.

For large CNN models such as ResNet50 and Inception-v3, optimized GPU kernel computation accounts for most of the computation time, which makes the performance difference among JANUS, TensorFlow, and TensorFlow Eager relatively small.

**Optimization effect.** Figure 7 analyzes the cause of the performance improvement of JANUS in detail. Converting the imperative program into the symbolic graph without any following optimizations (**BASE**) enabled up to 4.9x performance improvement compared to the imperative execution (**IMP**). It removes the Python interpreter and framework code overhead, which has the bigger effect when each graph operation is relatively smaller. Control flow unrolling (**+UNRL**) and type specialization (**+SPCN**) enable more aggressive compiler optimizations. On RNNs, **+UNRL** improved the performance of LSTM and LM by 2.09x and 1.04x, respectively. The control flow statements in CNNs, TreeNNs and DRLs could not be unrolled due to their dynamicity. **+SPCN** enabled some compiler optimizations and improved the throughput up to 18.3% in small neural networks. Finally, executing multiple operations in parallel (**+PARL**) improved the throughput up to 9.81x. Especially higher gain could be achieved for TreeNNs, since there exist many operations that could be executed in parallel in multiple independent tree nodes.

We have also measured the effect of assumption validation, but the effect was negligible (in the error range), since the AssertOps can be executed with the main neural network in parallel.

### 6.3.2 Scalability

Figure 8 shows the scalability of ResNet50, Inception-v3, LM, and PPO models on JANUS, TensorFlow, and TensorFlow Eager on the cluster with 36 GPUs (12 GPUs for LM, 6 GPUs for PPO). We measured the scale factor, which is defined as *Multi-GPU Throughput / (Single-GPU Throughput × Number of GPUs)*. JANUS achieves similar scalability (scale factor 0.77, 0.81, 0.18 each) as TensorFlow (0.81, 0.80, 0.18 each), but TensorFlow Eager does not scale well (0.24, 0.24, 0.14 each), due its inability to overlap computation and communication.

The performance difference between JANUS and TensorFlow becomes smaller when the synthetic dataset is used, since the input processing of TensorFlow is highly optimized. The slight difference in the scalability of ResNet50 comes from the under-optimized input pipeline of TensorFlow Eager, which JANUS also uses. Optimizing the input processing pipeline for JANUS will further reduce the performance difference between JANUS and TensorFlow. We leave this optimization as future work.

## 7 Conclusion

In this paper, we introduced JANUS, a system that achieves the performance of symbolic DL frameworks while maintaining the programmability of imperative DL frameworks. To achieve the performance of symbolic DL frameworks, JANUS converts imperative DL programs into static dataflow graphs by assuming that DL programs inherently have the static nature. To preserve the dynamic semantics of Python, JANUS generates and executes the graph speculatively, verifying the correctness of such assumptions at runtime. Our experiments showed that JANUS can execute various deep neural networks efficiently while retaining programmability of imperative programming.

## Acknowledgments

We thank our shepherd Srikanth Kandula and the anonymous reviewers for their insightful comments. This work was supported by Samsung Advanced Institute of Technology, the AWS Machine Learning Research Awards program, and Institute for Information & communications Technology Promotion(IITP) grant funded by the Korea government(MSIT) (No.2015-0-00221, Development of a Unified High-Performance Stack for Diverse Big Data Analytics).

## References

[1] Martín Abadi, Paul Barham, Jianmin Chen, Zhifeng Chen, Andy Davis, Jeffrey Dean, Matthieu Devin, Sanjay Ghemawat, Geoffrey Irving, Michael Isard, Manjunath Kudlur, Josh Levenberg, Rajat Monga, Sherry Moore, Derek G. Murray, Benoit Steiner, Paul Tucker, Vijay Vasudevan, Pete Warden, Martin Wicke, Yuan Yu, and Xiaoqiang Zheng. Tensorflow: A system for large-scale machine learning. In *OSDI*, 2016.

[2] Arvind and Rishiyur S. Nikhil. Executing a program on the mit tagged-token dataflow architecture. *IEEE Transactions on computers*, 39(3):300–318, March 1990.

[3] Jeff Bezanson, Alan Edelman, Stefan Karpinski, and Viral B. Shah. Julia: A fresh approach to numerical computing. *Siam Review*, 59(1):65–98, February 2017.

[4] Greg Brockman, Vicki Cheung, Ludwig Pettersson, Jonas Schneider, John Schulman, Jie Tang, and Wojciech Zaremba. Openai gym. *CoRR*, abs/1606.01540, 2016.

[5] Lars Buitinck, Gilles Louppe, Mathieu Blondel, Fabian Pedregosa, Andreas Mueller, Olivier Grisel, Vlad Niculae, Peter Prettenhofer, Alexandre Gramfort, Jaques Grobler, Robert Layton, Jake VanderPlas, Arnaud Joly, Brian Holt, and Gaël Varoquaux. API design for machine learning software: experiences from the scikit-learn project. In *LML Workshop at ECML PKDD*, 2013.

[6] Ciprian Chelba, Tomas Mikolov, Mike Schuster, Qi Ge, Thorsten Brants, Phillipp Koehn, and Tony Robinson. One billion word benchmark for measuring progress in statistical language modeling. Technical report, Google, 2013.

[7] Tianqi Chen, Mu Li, Yutian Li, Min Lin, Naiyan Wang, Minjie Wang, Tianjun Xiao, Bing Xu, Chiyuan Zhang, and Zheng Zhang. MXNet: A flexible and efficient machine learning library for heterogeneous distributed systems. In *Workshop on Machine Learning Systems in NIPS*, 2015.

[8] François Chollet et al. Keras. https://keras.io.

[9] David E Culler. Dataflow architectures. *Annual review of computer science*, 1(1):225–253, June 1986.

[10] Jeffrey Dean and Sanjay Ghemawat. Mapreduce: simplified data processing on large clusters. In *OSDI*, 2004.

[11] Jack B Dennis and David P Misunas. A preliminary architecture for a basic data-flow processor. *ACM SIGARCH Computer Architecture News*, 3(4):126–132, December 1974.

[12] Facebook. Caffe2. https://caffe2.ai.

[13] Roy Frostig, Matthew James Johnson, and Chris Leary. Compiling machine learning programs via high-level tracing. In *SysML*, 2018.

[14] Ian Goodfellow, Jean Pouget-Abadie, Mehdi Mirza, Bing Xu, David Warde-Farley, Sherjil Ozair, Aaron Courville, and Yoshua Bengio. Generative adversarial nets. In *NIPS*, 2014.

[15] William Gropp, Ewing Lusk, and Anthony Skjellum. *Using MPI: portable parallel programming with the message-passing interface*, volume 1. MIT press, 1999.

[16] Kaiming He, Xiangyu Zhang, Shaoqing Ren, and Jian Sun. Deep residual learning for image recognition. In *CVPR*, 2016.

[17] Sergey Ioffe and Christian Szegedy. Batch normalization: Accelerating deep network training by reducing internal covariate shift. In *ICML*, 2015.

[18] Michael Isard, Mihai Budiu, Yuan Yu, Andrew Birrell, and Dennis Fetterly. Dryad: distributed data-parallel programs from sequential building blocks. In *EuroSys*, 2007.

[19] Phillip Isola, Jun-Yan Zhu, Tinghui Zhou, and Alexei A. Efros. Image-to-image translation with conditional adversarial networks. In *CVPR*, 2017.

[20] Eunji Jeong, Joo Seong Jeong, Soojeong Kim, Gyeong-In Yu, and Byeong-Gon Chun. Improving the expressiveness of deep learning frameworks with recursion. In *EuroSys*, 2018.

[21] Rafal Józefowicz, Oriol Vinyals, Mike Schuster, Noam Shazeer, and Yonghui Wu. Exploring the limits of language modeling. *CoRR*, abs/1602.02410, 2016.

[22] Yann LeCun, Leon Buttou, Yoshua Bengio, and Patrick Haffner. Gradient-based learning applied to document recognition. *Proceedings of the IEEE*, 86(11):2278–2324, November 1998.

[23] Yann LeCun and Corinna Cortes. The mnist database of handwritten digits. http://yann.lecun.com/exdb/mnist/.

[24] Moshe Looks, Marcello Herreshoff, DeLesley Hutchins, and Peter Norvig. Deep learning with dynamic computation graphs. In *ICLR*, 2017.

[25] Azalia Mirhoseini, Hieu Pham, Quoc V. Le, Benoit Steiner, Rasmus Larsen, Yuefeng Zhou, Naveen Kumar, Mohammad Norouzi, Samy Bengio, and Jeff Dean. Device placement optimization with reinforcement learning. *CoRR*, abs/1706.04972, 2017.

[26] Volodymyr Mnih, Adrià Puigdomènech Badia, Mehdi Mirza, Alex Graves, Timothy P. Lillicrap, Tim Harley, David Silver, and Koray Kavukcuoglu. Asynchronous methods for deep reinforcement learning. In *ICML*, 2016.

[27] Dan Moldovan, James M. Decker, Fei Wang, Andrew A. Johnson, Brian K. Lee, Zachary Nado, D. Sculley, Tiark Rompf, and Alexander B. Wiltschko. Autograph: Imperative-style coding with graph-based performance. *CoRR*, abs/1810.08061, 2018.

[28] MXNet. Deep Learning Programming Style. https://mxnet.incubator.apache.org/architecture/program_model.html.

[29] MXNet Developers. Gluon. http://gluon.mxnet.io/.

[30] Stefan C. Müller, Gustavo Alonso, and Adam Amara André Csillaghy. Pydron: Semi-automatic parallelization for multi-core and the cloud. In *OSDI*, 2014.

[31] Graham Neubig, Chris Dyer, Yoav Goldberg, Austin Matthews, Waleed Ammar, Antonios Anastasopoulos, Miguel Ballesteros, David Chiang, Daniel Clothiaux, Trevor Cohn, Kevin Duh, Manaal Faruqui, Cynthia Gan, Dan Garrette, Yangfeng Ji, Lingpeng Kong, Adhiguna Kuncoro, Gaurav Kumar, Chaitanya Malaviya, Paul Michel, Yusuke Oda, Matthew Richardson, Naomi Saphra, Swabha Swayamdipta, and Pengcheng Yin. Dynet: The dynamic neural network toolkit. *CoRR*, abs/1701.03980, 2017.

[32] Adam Paszke, Sam Gross, Soumith Chintala, Gregory Chanan, Edward Yang, Zachary DeVito, Zeming Lin, Alban Desmaison, Luca Antiga, and Adam Lerer. Automatic differentiation in pytorch. In *Autodiff Workshop in NIPS*, 2017.

[33] Python Software Foundation. Python programming language. https://www.python.org/.

[34] Olga Russakovsky, Jia Deng, Hao Su, Jonathan Krause, Sanjeev Satheesh, Sean Ma, Zhiheng Huang, Andrej Karpathy, Aditya Khosla, Michael Bernstein, et al. Imagenet large scale visual recognition challenge. *International Journal of Computer Vision*, 115(3):211–252, December 2015.

[35] John Schulman, Filip Wolski, Prafulla Dhariwal, Alec Radford, and Oleg Klimov. Proximal policy optimization algorithms. *CoRR*, abs/1707.06347, 2017.

[36] Alexander Sergeev and Mike Del Balso. Horovod: fast and easy distributed deep learning in tensorflow. *CoRR*, abs/1802.05799, 2018.

[37] Richard Socher, Cliff Chiung-Yu Lin, Andrew Y. Ng, and Christopher D. Manning. Parsing natural scenes and natural language with recursive neural networks. In *ICML*, 2011.

[38] Richard Socher, Alex Perelygin, Jean Wu, Jason Chuang, Christopher D Manning, Andrew Ng, and Christopher Potts. Recursive deep models for semantic compositionality over a sentiment treebank. In *EMNLP*, 2013.

[39] Christian Szegedy, Vincent Vanhoucke, Sergey Ioffe, Jon Shlens, and Zbigniew Wojna. Rethinking the inception architecture for computer vision. In *CVPR*, 2016.

[40] Kai Sheng Tai, Richard Socher, and Christopher D Manning. Improved semantic representations from tree-structured long short-term memory networks. In *ACL*, 2015.

[41] TensorFlow. Eager Execution. https://www.tensorflow.org/programmers_guide/eager.

[42] TensorFlow. GAN with TensorFlow eager execution. https://github.com/tensorflow/tensorflow/tree/master/tensorflow/contrib/eager/python/examples/gan.

[43] TensorFlow. Swift for TensorFlow. https://github.com/tensorflow/swift.

[44] TensorFlow. tf.contrib.eager.defun. https://www.tensorflow.org/versions/master/api_docs/python/tf/contrib/eager/defun.

[45] TensorFlow. XLA Overview. https://www.tensorflow.org/performance/xla/.

[46] Theano Development Team. Theano: A python framework for fast computation of mathematical expressions. *CoRR*, abs/1605.02688, 2016.

[47] Radim Tyleček and Radim Šára. Spatial pattern templates for recognition of objects with regular structure. In *GCPR*, 2013.

[48] Stéfan Van Der Walt, S. Chris Colbert, and Gael Varoquaux. The numpy array: A structure for efficient numerical computation. *Computing in Science Engineering*, 13(2):22–30, March 2011.

[49] Dong Yu, Adam Eversole, Mike Seltzer, Kaisheng Yao, Zhiheng Huang, Brian Guenter, Oleksii Kuchaiev, Yu Zhang, Frank Seide, Huaming Wang, et al. An introduction to computational networks and the computational network toolkit. *Microsoft Technical Report MSR-TR-2014–112*, 2014.

[50] Yuan Yu, Martín Abadi, Paul Barham, Eugene Brevdo, Mike Burrows, Andy Davis, Jeff Dean, Sanjay Ghemawat, Tim Harley, Peter Hawkins, Michael Isard, Manjunath Kudlur, Rajat Monga, Derek Murray, and Xiaoqiang Zheng. Dynamic control flow in large-scale machine learning. In *EuroSys*, 2018.

[51] Wojciech Zaremba, Ilya Sutskever, and Oriol Vinyals. Recurrent neural network regularization. *CoRR*, abs/1409.2329, 2014.

# Appendix

## A  Python Syntax Coverage

Table 4 describes the entire set of opcode in the CPython [33] 3.5.2 interpreter, and maps them to the sections which describe the corresponding graph generation rules. Python programs whose opcodes are mapped to Section 4.3 can only be executed on the imperative executor, and the others can be executed on the graph executor. Python features that are not covered in previous sections are briefly discussed in the rest of this section.

**Exceptions.** A Python raise statement can be represented as an AssertOp in the dataflow graph. When the AssertOp for an exception aborts the graph execution, the fallback occurs, and the actual, Python-style exception can be safely raised on the imperative executor. Under the same principle, for try-except-finally statements, only the try-finally part is converted into the graph elements, and the except part is simply not converted, since the exception will never be caught by the symbolic graph. By avoiding exception handling inside the symbolic graph, we can protect users from having to debug through symbolic graph execution traces, which are relatively more complicated than imperative execution traces.

**Context manager.** Since exception handling always occurs on the imperative executor as described in the previous paragraph, the with statement can be converted into the simple function calls to \_\_enter\_\_ and \_\_exit\_\_ of the corresponding context manager object.

| Opcode | Num | Description | Section Ref. |
|---|---|---|---|
| POP_TOP, ROT_TWO, ROT_THREE, DUP_TOP, DUP_TOP_TWO, NOP, EXTENDED_ARG | 7 | stack manipulation | No conversion is necessary |
| LOAD_CONST | 1 | constant | Section 4.1 |
| UNARY_INVERT, UNARY_NEGATIVE, UNARY_NOT, UNARY_POSITIVE, BINARY_ADD, BINARY_AND, BINARY_FLOOR_DIVIDE, BINARY_LSHIFT, BINARY_MATRIX_MULTIPLY, BINARY_MODULO, BINARY_MULTIPLY, BINARY_OR, BINARY_POWER, BINARY_RSHIFT, BINARY_SUBTRACT, BINARY_TRUE_DIVIDE, BINARY_XOR, INPLACE_ADD, INPLACE_AND, INPLACE_FLOOR_DIVIDE, INPLACE_LSHIFT, INPLACE_MATRIX_MULTIPLY, INPLACE_MODULO, INPLACE_MULTIPLY, INPLACE_OR, INPLACE_POWER, INPLACE_RSHIFT, INPLACE_SUBTRACT, INPLACE_TRUE_DIVIDE, INPLACE_XOR, COMPARE_OP | 31 | mathematical operators | Section 4.1 |
| LOAD_FAST, STORE_FAST, DELETE_FAST, UNPACK_SEQUENCE, UNPACK_EX | 5 | local variables | Section 4.1 |
| JUMP_ABSOLUTE, JUMP_FORWARD, JUMP_IF_FALSE_OR_POP, JUMP_IF_TRUE_OR_POP, POP_JUMP_IF_FALSE, POP_JUMP_IF_TRUE, POP_BLOCK, GET_ITER, FOR_ITER, BREAK_LOOP, CONTINUE_LOOP, SETUP_LOOP | 12 | dynamic control flow | Section 4.2.1 |
| CALL_FUNCTION, CALL_FUNCTION_KW, CALL_FUNCTION_VAR, CALL_FUNCTION_VAR_KW, RETURN_VALUE, MAKE_FUNCTION | 6 | function call | Section 4.2.1, Section 4.3.1 |
| LOAD_ATTR, STORE_ATTR, DELETE_ATTR | 3 | arbitrary object | Section 4.2.2, Section 4.2.3 |
| BUILD_LIST, BUILD_LIST_UNPACK, LIST_APPEND, BUILD_MAP, BUILD_MAP_UNPACK, BUILD_MAP_UNPACK_WITH_CALL, MAP_ADD, BUILD_SET, BUILD_SET_UNPACK, SET_ADD, BUILD_SLICE, BUILD_TUPLE, BUILD_TUPLE_UNPACK, BINARY_SUBSCR, STORE_SUBSCR, DELETE_SUBSCR | 16 | list, set, map | Section 4.2.2, Section 4.2.3 |
| LOAD_GLOBAL, LOAD_DEREF, LOAD_NAME, STORE_GLOBAL, STORE_DEREF, STORE_NAME, DELETE_GLOBAL, DELETE_DEREF, DELETE_NAME, LOAD_CLOSURE, MAKE_CLOSURE | 11 | non-local variables | Section 4.2.3 |
| POP_EXCEPT, SETUP_EXCEPT, SETUP_FINALLY, RAISE_VARARGS, END_FINALLY | 5 | exception handling | Appendix A |
| SETUP_WITH, WITH_CLEANUP_FINISH, WITH_CLEANUP_START | 3 | `with` | Appendix A |
| YIELD_FROM, YIELD_VALUE, GET_YIELD_FROM_ITER | 3 | `yield` | Section 4.3.2 |
| IMPORT_FROM, IMPORT_NAME, IMPORT_STAR | 3 | in-line `import` | Section 4.3.2 |
| LOAD_BUILD_CLASS, LOAD_CLASSDEREF | 2 | in-line class definition | Section 4.3.2 |
| GET_AITER, GET_ANEXT, GET_AWAITABLE, BEFORE_ASYNC_WITH, SETUP_ASYNC_WITH | 5 | coroutine | Section 4.3.2 |
| Total | 113 | | |

Table 4: The mapping of the full list of CPython opcode and the corresponding sections.